# Maximal Spectral Efficiency of OFDM with Index Modulation under Polynomial Space Complexity

Saulo Queiroz*†, Wesley Silva*, João P. Vilela†, Edmundo Monteiro†
*Academic Department of Informatics (DAINF)
Federal University of Technology (UTFPR), Ponta Grossa, PR, Brazil.
sauloqueiroz@utfpr.edu.br, wesley.1999@alunos.utfpr.edu.br
†CISUC and Department of Informatics Engineering
University of Coimbra, Portugal
{saulo, jpvilela, edmundo}@dei.uc.pt

*Abstract*—In this letter, we demonstrate a mapper that enables all waveforms of OFDM with Index Modulation (OFDM-IM) while preserving polynomial time and space computational complexities. Enabling *all OFDM-IM waveforms* maximizes the spectral efficiency (SE) gain over the classic OFDM but, as far as we know, the computational overhead of the resulting mapper remains conjectured as prohibitive across the OFDM-IM literature. We show that the largest number of binomial coefficient calculations performed by the original OFDM-IM mapper is polynomial on the number of subcarriers, even under the setup that maximizes the SE gain over OFDM. Also, such coefficients match the entries of the so-called Pascal's triangle (PT). Thus, by assisting the OFDM-IM mapper with a PT table, we show that the maximum SE gain over OFDM can be achieved under polynomial (rather than exponential) time and space complexities.

## I. INTRODUCTION

Look-up table (LUT) is a fundamental technique for the efficient implementation of OFDM mappers. In an $N$-subcarrier OFDM system, the mapper translates $N \log_2 M$ bits into $N$ complex baseband samples chosen from an $M$-ary constellation diagram. To achieve better spectral efficiency (SE), novel physical layer techniques are expected to map more bits in the same amount of spectrum. In this context, index modulation (IM) has gained increasing attention in the literature [1], [2]. In the OFDM with IM (OFDM-IM) reference design [3], only $k \leq N$ subcarriers of the symbol are active, which enables $C(N,k) = \binom{N}{k} = N!/(k!(N-k)!)$ waveforms. Of these, OFDM-IM employs $2^{\lfloor \log_2 C(N,k) \rfloor}$ to map $p_1 = \lfloor \log_2 C(N,k) \rfloor$ bits. Further, by employing an $M$-ary constellation for the active subcarriers, more $p_2 = k \log_2 M$ bits can be modulated, yielding a total of $m = p_1 + p_2$ bits per OFDM-IM symbol.

This work is supported by the European Regional Development Fund (FEDER), through the Regional Operational Programme of Lisbon (POR LISBOA 2020) and the Competitiveness and Internationalization Operational Programme (COMPETE 2020) of the Portugal 2020 framework [Project 5G with Nr. 024539 (POCI-01-0247-FEDER-024539)], and by the CONQUEST project - CMU/ECE/0030/2017 Carrier AggregatiON between Licensed Exclusive and Licensed Shared Access FreQUEncy BandS in HeTerogeneous Networks with Small Cells.

The selection of the $k$ indexes to activate in the symbol is determined from the $p_1$-bit sequence in a DSP step called Index Selector (IxS). The IxS computation can be as efficient as OFDM's mapping if implemented as a $2^{p_1}$-entry LUT. However, it is consensual in the OFDM-IM literature that an IxS LUT should be employed only for small $N$ because the required storage is practically infeasible otherwise. For large $N$, the OFDM-IM mapper employs an online IxS algorithm. However, when the setup that maximizes the OFDM-IM SE gain over OFDM is chosen, i.e., $M=2$ and $k=N/2$ ($N$ is even) [4] (we refer to as the "ideal setup"), the original IxS algorithm runs in $O(N^2)$ steps and becomes the most complex DSP block of the OFDM-IM transmitter [5]. According to [3], the IxS computational overhead is the reason why the IM technique is restricted to small parts of the symbol instead of being applied to all $N$ subcarriers. This approach, called subblock partitioning (SP), alleviates the IxS complexity at the penalty of preventing the SE maximization. Novel constellation designs improve the OFDM-IM SE by modulating extra data on the IM subcarriers e.g. [6], [7]. Yet, these proposals employ SP to mitigate the IxS complexity, which prevents them to reach their respective maximum theoretical number of IM waveforms.

In this letter, we study the asymptotic trade-off between spectral and computational resources (space and time)[1] in the OFDM-IM mapper design. We demonstrate that OFDM-IM requires a $\Theta(2^N/\sqrt{N})$-entry LUT to reduce the IxS time complexity from $O(N^2)$ to $O(N)$ and to keep its maximum SE gain over OFDM. Then, we show how both the time complexity reduction and the SE maximization can be met by replacing the traditional LUT with the so-called Pascal's triangle (PT). To the best of our knowledge, this is the first demonstration of an OFDM-IM mapper running at the same asymptotic complexity of a $2^{p_1}$-entry LUT requiring neither extra exponential space complexity nor sacrificing the ideal OFDM-IM setup. Our results demonstrate the computational feasibility of the ideal (SP-free) OFDM-IM mapper, which has been conjectured as computationally infeasible in the literature [2].

---

[1] In this work, we refer to "time" and "space" as the asymptotic number of computational steps and table entries required by an $N$-subcarrier mapper, respectively.

**Algorithm 1** OFDM-IM index selector algorithm.
1: $\{X, k,$ and $N$ are input parameters. Array $c$ is returned$\}$;
2: $largestCandidate \leftarrow N - 1$;
3: **for** $i = k$ **downto** $1$ **do**
4: $\quad c_i \leftarrow largestCandidate$;
5: $\quad \{C(c_i, i)$ is computed in $O(i)$ from $\prod_{j=1}^{i} \frac{c_i - j + 1}{j}\}$
6: $\quad$ **while** $C(c_i, i) > X$ **do**
7: $\quad\quad c_i \leftarrow c_i - 1$;
8: $\quad$ **end while**
9: $\quad X \leftarrow X - C(c_i, i)$;
10: $\quad largestCandidate \leftarrow c_i - 1$;
11: **end for**

## II. OFDM-IM Mapper Spectro-Computational Trade-Offs

In this section, we review the IxS algorithm (Subsection II-A) and derive the asymptotic number of entries of the OFDM-IM LUT (Subsection II-B). In Subsection II-C, we present a mapper that enables all $2^{p_1}$ OFDM-IM waveforms while keeping polynomial time and space complexities. Then, in Subsection II-D, we analyze the throughput of our mapper.

### A. Spectro-Time Trade-Off

The IxS algorithm of OFDM-IM (Alg. 1) is based on the combinatorial number system. This system tells us that, for every integer $X \in [0, C(N, k) - 1]$, there exist $k$ integers $c_k > \cdots > c_2 > c_1 \geq 0$ such that $X = \sum_{i=1}^{k} C(c_i, i)$ and $c_k < N$. For OFDM-IM, $X$ is the decimal representation of the $p_1$-bit input and the coefficients $c_k, \cdots, c_1$ are the indexes of the active subcarriers. Given $X$, $k$, and $N$ as input, the IxS algorithm computes the active subcarriers from $c_k$ to $c_1$. In its first round, the IxS algorithm determines the value for $c_k$. To this end, it assigns $c_k$ with the largest candidate value $N - 1$ (line 2) and checks whether $C(c_k, k) \leq X$ holds (line 6). If this logic test fails, the inner loop keeps decrementing $c_k$ (so, recalculating $C(c_k, k)$) until the test passes. When this happens, $X$ is decremented by $C(c_k, k)$ and the largest candidate value available for the next coefficient $c_{k-1}$ is $c_k - 1$. This process repeats for all remainder coefficients. Since the efficient calculation of a single binomial coefficient $C(c_i, i)$ takes $O(i)$ iterations with the multiplicative formula $\prod_{j=1}^{i}(c_i + 1 - j)/j$, all $k$ coefficients $C(c_k, k), \cdots, C(c_1, 1)$ computed by the IxS algorithm take a total of $k + \cdots + 2 + 1 = k(k+1)/2 = O(k^2)$ iterations. As explained in [5], this complexity becomes $O(N^2)$ if SE maximizes, surpassing IFFT as the most complex block of the OFDM-IM transmitter.

### B. Spectro-Storage Trade-Off

A LUT for all $2^{p_1}$ entries is pointed as an alternative to the IxS algorithm for relatively "small" $N$ [3]. In the Lemma 2, we show that a $2^{p_1}$-entry LUT provides OFDM-IM with $O(N)$ time complexity, the same as the classical OFDM mapper (Corollary 1). This time complexity holds even under the ideal SE setup of OFDM-IM. However, the LUT size becomes prohibitive as $N$ grows. In Lemma 3 we demonstrate that the asymptotic number of entries of the OFDM-IM LUT under the ideal OFDM-IM setup is $\Theta(2^N/\sqrt{N})$. Both Lemmas 2 and 3 rely on the fact that $p_1$ can be asymptotically approximated by $N - \log_2 \sqrt{N}$ under the ideal OFDM-IM setup (Lemma 1).

**Lemma 1** (Maximum Number $p_1$ of Index Modulation Bits). *The maximum number of index modulated bits $p_1$ approaches $N - \log_2 \sqrt{N}$ for arbitrarily large $N$.*

*Proof.* By definition, $p_1 = \lfloor \log_2 C(N, k) \rfloor$. If the maximum SE gain of OFDM-IM over OFDM is allowed, $C(N, k)$ becomes the so-called central binomial coefficient $C(N, N/2) = \Theta(2^N N^{-0.5})$ [5]. From this, it follows that $p_1$ approaches $\log_2(2^N N^{-0.5}) = N - \log_2 \sqrt{N}$ as $N \to \infty$. $\square$

**Lemma 2** (LUT-Based OFDM-IM Mapper Complexity). *A $2^{p_1}$-entry LUT enables the OFDM-IM IxS to run in $O(N)$.*

*Proof.* Let $0 \leq X \leq 2^{p_1} - 1$ be the decimal representation of the $p_1$-bit input given to the IxS DSP block. If the IxS is a $2^{p_1}$-entry LUT indexed from 0 to $2^{p_1} - 1$, then the $k$-list of active indexes corresponding to $X$ is stored in the $X$-th entry of the table. Since LUTs are based on random access storage technology, any data can be retrieved in $O(1)$ time *after* the LUT index is read (which is $X$, in this case). Therefore, the time complexity of a LUT-based IxS is determined by the time to read $X$, which is $O(p_1) = O(\log_2 C(N, N/2)) = O(N - \log_2 \sqrt{N}) = O(N)$ (Lemma 1). Also, since the modulation of the $k = N/2$ active subcarriers follows as in the classic OFDM for $M = 2$, more $O(N/2)$ computations are required. Thus, if the IxS is implemented as a $2^{p_1}$-entry LUT, the overall OFDM-IM mapper runs in $O(N) + O(N/2) = O(N)$ time. $\square$

**Corollary 1** (OFDM-IM Mapper As Fast As OFDM's). *A $2^{p_1}$-entry Index Selector (IxS) LUT enables the OFDM-IM mapper to run as efficiently as OFDM's.*

*Proof.* The OFDM mapper reads $N$ independent groups of $\log_2 M$ bits to produce $N$ complex baseband samples. Therefore, its total number of computational steps is $N \log_2 M$, which reduces to $N = \Theta(N)$ steps under the "ideal OFDM-IM setup" (i.e., $M = 2$). Thus, according to the definition of the asymptotic notation, an OFDM-IM mapper runs at the same time complexity of an OFDM mapper if its total number of computational steps is $\kappa N$ (for some constant $\kappa > 0$ and $N \to \infty$). As shown in Lemma 2, this is the case of a $2^{p_1}$-entry LUT mapper. Therefore, by consuming $\Theta(2^{p_1})$ of space, the time complexity of the OFDM-IM mapper becomes as efficient as the time complexity of the OFDM mapper. $\square$

**Lemma 3** (OFDM-IM LUT Size Under Maximal SE). *Under the ideal OFDM-IM setup, a LUT-based OFDM-IM mapper requires $\Theta(2^N/\sqrt{N})$ entries.*

*Proof.* A LUT-based OFDM-IM mapper has one entry per each one of all possible $2^{p_1} = 2^{\lfloor \log_2 C(N,k) \rfloor}$ symbol waveforms. Since $p_1$ approaches $N - \log_2 \sqrt{N}$ as $N$ grows (Lemma 1), the number of LUT entries approaches $\Theta(2^{N - \log_2 \sqrt{N}}) = \Theta(2^N/\sqrt{N})$. $\square$

### C. Spectral Efficiency Maximization under Polynomial Space

In Subsection II-B, we formalize the time and space trade-off of the original OFDM-IM mapper. The mapper needs an exponential amount of space to support all $2^{p_1}$ waveforms at the same asymptotic time of OFDM. We note that this trade-off can be improved if the OFDM-IM mapper is assisted by the so-called Pascal's triangle (PT) instead of being implemented as a $2^{p_1}$-entry LUT. The proposed mapper is illustrated in Fig. 1. It consists of the original OFDM-IM mapper set to a single subblock and having the IxS algorithm assisted by a PT table. The PT table can be viewed as an $N \times k$ matrix that stores the result of $C(c_i, i)$ in row $c_i$ and column $i$ (Table I). This way, the $O(i)$ iterations required to compute a single binomial coefficient $C(c_i, i)$ is replaced by a single query to the PT table. Therefore, the $O(k^2)$ iterations performed by the IxS algorithm to compute the $k$ binomial coefficients $C(c_k, k), \cdots, C(c_1, 1)$ (as explained in Subsection II-A) can be replaced by $O(k)$ queries to the PT table.

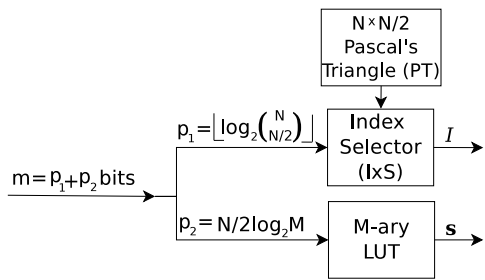

Fig. 1. Proposed OFDM-IM mapper. Under maximal spectral efficiency, the value of any binomial coefficient $C(c_i, i)$ required in line 6 of the IxS block (Algorithm 1) matches an entry of the PT table shown in Table I. By querying the table for each $C(c_i, i)$ instead of calculating them from scratch, the mapper achieves the same time complexity of the $\Theta(2^N/\sqrt{N})$-entry OFDM-IM look-up table storing only $\Theta(N^2)$ entries.

Note that the time complexity improvement achieved by the PT table does not change the $k$ binomial coefficients selected by the IxS algorithm. Hence, both the vector of active indexes and the vector of complex baseband samples (denoted as $I$ and $\mathbf{s}$ in Fig. 1, respectively) remain the same as in the original OFDM-IM mapper. In Lemma 4, we show that the number of binomial coefficient entries of the PT table grows polynomially on $N$ even if all $2^{p_1}$ OFDM-IM waveforms are enabled. Then, in Thm. 1, we show that the PT table can enable all $2^{p_1}$ OFDM-IM waveforms at the same time complexity of the OFDM mapper.

**Lemma 4** (Binomial Coefficients under Maximal SE). *Under the ideal SE setup, the OFDM-IM Index Selector algorithm computes $O(N^2)$ distinct binomial coefficients. Thus, a $\Theta(N^2)$-entry PT table can be employed to reduce the IxS time complexity from $O(N^2)$ to $O(N)$.*

*Proof.* Under the ideal OFDM-IM setup, the variables $c_i$ and $i$ of the IxS algorithm (Alg. 1, Subsec. II-A) decrease by 1 starting from $N-1$ and $k = N/2$, respectively. Hence, the algorithm needs to compute no binomial coefficient other than $C(c_i, i)$, $0 \le c_i \le N-1$ and $1 \le i \le N/2$. Therefore, the $\Theta(N^2)$-entry PT of Table I enables any binomial coefficient required by the IxS algorithm to be returned in $O(1)$ time. Thus, the inner loop of the IxS algorithm (Alg. 1) reduces from $O(k) \times O(i)$ to $O(k) \times O(1)$, yielding to an overall complexity of $O(k) = O(N)$ in the ideal OFDM-IM setup. □

TABLE I
PASCAL'S TRIANGLE OF THE PROPOSED OFDM-IM MAPPER (FIG. 1).

| $c_i \mid i$ | 1 | 2 | 3 | $\cdots$ | $N/2$ |
|---|---|---|---|---|---|
| 0 | 0 | 0 | 0 | $\cdots$ | 0 |
| 1 | 1 | 0 | 0 | $\cdots$ | 0 |
| 2 | 2 | 1 | 0 | $\cdots$ | 0 |
| 3 | 3 | 3 | 1 | $\cdots$ | 0 |
| 4 | 4 | 6 | 4 | $\cdots$ | 0 |
| $\vdots$ | $\vdots$ | $\vdots$ | $\vdots$ | $\ddots$ | $\vdots$ |
| $N-1$ | $\binom{N-1}{1}$ | $\binom{N-1}{2}$ | $\binom{N-1}{3}$ | $\cdots$ | $\binom{N-1}{N/2}$ |

**Theorem 1** (OFDM-IM Mapper under Polynomial Space). *All $2^{p_1}$ OFDM-IM waveforms can be mapped at the same asymptotic time of an OFDM mapper at the expense of polynomial space complexity.*

*Proof.* From Lemma 4, a PT table storing $\Theta(N^2)$ binomial coefficients enables the IxS algorithm to run in $O(N)$ time. This is the same asymptotic number of steps performed by the OFDM mapper (Corollary 1). To achieve such time complexity keeping the ideal setup, a traditional LUT-based OFDM-IM mapper requires $\Theta(2^N/\sqrt{N})$ entries (Lemma 3). Thus, by replacing a LUT with an IxS algorithm assisted by the PT table, one enables the OFDM-IM mapper to achieve its ideal SE setup in $O(N)$ time at the expense of polynomial (rather than exponential) space. □

The PT table dates back from ancient times, even before Blaise Pascal[2]. Thus, the improvement it provides for the calculation of binomial coefficients is not a novelty for the field of combinatorial algorithms. Nonetheless, how this result turns out to affect the comparative SE performance of OFDM-IM and OFDM is beyond the scope of that literature.

### D. Mapper Throughput Analysis

In [5] the authors propose the spectro-computational efficiency (SCE) metric to formalize the OFDM-IM trade-off between SE and computational complexity. Based on SCE, in Def. 1 we present the condition for the scalability of the OFDM-IM mapper throughput. In Thm. 2, we show that the throughput of our mapper scales under polynomial space complexity.

**Definition 1** (Scalable Throughput [5]). *Let $T(N)$ be the computational complexity to map $m(N)$ bits into an $N$-subcarrier symbol and $m(N)/T(N)$ be the throughput of the mapper. Then, the throughput of the mapper is not scalable unless ineq. 1 does hold.*

$$\lim_{N \to \infty} \frac{m(N)}{T(N)} > 0 \quad (1)$$

**Theorem 2** (Scalable OFDM-IM Mapper Throughput under Polynomial Space). *The throughput $m(N)/T(N)$ (Def. 1) of our mapper does scale under polynomial space complexity.*

[2]for references before Blaise Pascal please, refer to https://en.wikipedia.org/wiki/Pascal's_triangle#History.

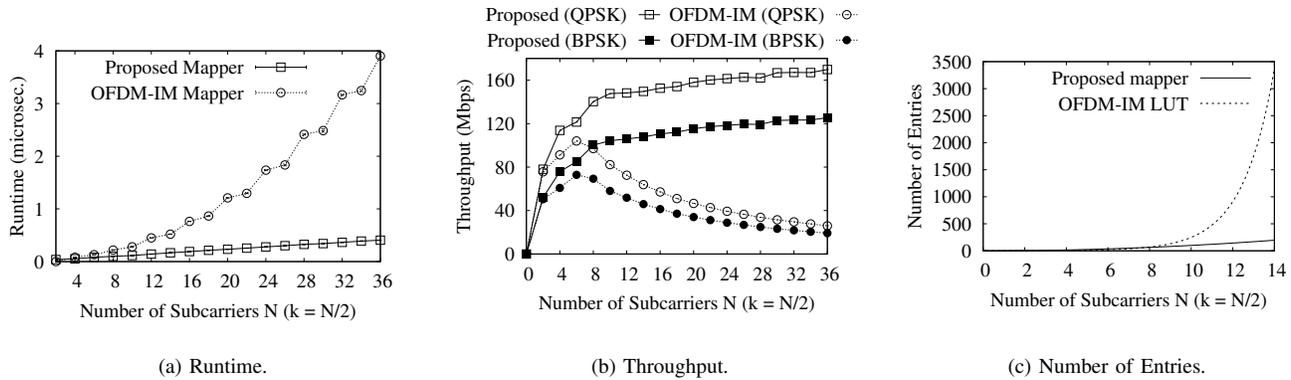

Fig. 2. Proposed mapper vs. OFDM-IM index selector mapper under the maximal spectral efficiency: runtime (a), throughput (b) and number of table entries compared to the LUT-based OFDM-IM mapper (c).

*Proof.* Under its ideal SE, the proposed OFDM-IM mapper run in $T(N) = O(N)$ time at the expense of a $\Theta(N^2)$-entry LUT (Thm. 1). Hence, as $N \to \infty$, $T(N)$ is bounded by $\kappa \cdot N$, for some constant $\kappa > 0$. Similarly, $\lfloor \log_2 C(N, N/2) \rfloor$ approaches $N - \log_2 \sqrt{N}$ as $N \to \infty$ (Lemma 1). Thus, the throughput of our mapper does not nullify over $N$ since it results in a constant $\kappa' > 0$ (Ineq. 2). Therefore, the throughput of our mapper does scale under polynomial space complexity.

$$\lim_{N \to \infty} \frac{N/2 + N - \log_2 \sqrt{N}}{\kappa \cdot N} = \kappa' \quad > \quad 0 \qquad (2)$$

□

## III. NUMERICAL PERFORMANCE

In this section, we present a software-based case study for our theoretical findings. We assess the runtime $T(N)$ and the throughput $m(N)/T(N)$ (Def. 1) of our proposed mapper (Fig. 1) and the original OFDM-IM mapper. We average $T(N)$ adopting the same infinite-horizon methodology of [5]. We compute the number of bits $m(N)$ assuming $k = N/2$ for $M = 2$ (BPSK) and $M = 4$ (QPSK). In Figs. 2a and 2b, we plot the runtime and the throughput of the mappers for different values of $N$, respectively. As predicted, in the ideal setup, the runtime growth of the original IxS algorithm (shown in Fig. 2a) nullifies the throughput of the OFDM-IM mapper as $N$ grows (Fig. 2b). This happens even if $M$ increases (e.g., from BPSK to QPSK) because the IxS complexity depends only on $N$ and $k$. If the PT table assists the IxS algorithm (proposed mapper), the time complexity improves from $O(N^2)$ to $O(N)$ (Fig. 2a), and the resulting overhead does not nullify the throughput (Fig. 2b). To achieve such similar scalability, the current literature [1]–[4], [6] needs a LUT whose space becomes prohibitive even for relatively small $N$, as shown in Fig. 2c. Although non-exhaustive, this case study confirms our theoretical predictions, ensuring the OFDM-IM mapper does not need SP to enable all its $2^{\lfloor \log_2 C(N,N/2) \rfloor}$ waveforms.

## IV. CONCLUSION AND FUTURE WORK

In this letter, we show that the time complexity of a Look-up table (LUT)-based OFDM-IM mapper can be achieved under polynomial (rather than exponential) space complexity. By assisting the OFDM-IM index selector algorithm with the so-called Pascal's triangle, we demonstrate for the first time the non-necessity of compromise approaches that prevail in the OFDM-IM literature such as subblock partitioning (SP) [1]–[4], [6] and adoption of few active subcarriers [8]. In such approaches, the *maximal* SE is sacrificed to attenuate the mapping computational complexity. Therefore, *our mapper represents a step towards the deactivation of SP*. Future work may improve our proposal by achieving the same space of the OFDM mapper (i.e., $O(N)$ rather than $\Theta(N^2)$) or by considering other relevant performance indicators e.g., bit-error rate [9]. Also, our mapper can be adapted to spatial IM systems [10] and inspire the deactivation of SP in future versions of OFDM-IM and variants thereof [6], [7].